# Multi-Channel Silicon-Organic Hybrid PICs for 200G/λ and 400G/λ PAM4 Transmission

Lewis E. Johnson[(1*)], Ari Novack[(2)], Scott R. Hammond[(1)], Meisam Bahadori,[(2)] Kevin M. O'Malley[(1)], Delwin L. Elder[(1)], Matthew Streshinsky[(2)], Brad Booth[(1)]

[(1)] NLM Photonics, 4000 Mason Road, Suite 300, Seattle WA 98195, USA
[(2)] Enosemi Inc, 1107 NE 45th Street, Suite 345, Seattle WA 98105, USA (acquired by AMD)
[(*)] Corresponding author, lewisj@nlmphotonics.com

**Abstract** *We demonstrate open-eye 224G PAM4 transmission in a 1.6T-DR8 PIC implementing low-$V_\pi$ silicon-organic hybrid modulators ($V_\pi L$ < 0.5 V-mm) with >80 3 dB GHz bandwidth and a variant capable of 400G/λ (> 110 GHz) for 3.2T-DR8. Both PIC variants use commercial crosslinkable organic electro-optic materials. ©2025 The Author(s)*

## Introduction

Achieving efficient data transmission at 200G/λ and beyond requires highly efficient electro-optic modulators; high efficiency is particularly crucial for linear pluggable optics (LPO) and co-packaged optics (CPO). As existing silicon-photonic pn-junction based modulators have been increasingly pressured, a number of alternatives[1, 2] have been examined by the industry, including indium phosphide (InP), thin-film lithium niobate (TFLN), barium titanate (BTO), and plasmonics.[3]

Silicon-organic hybrid (SOH) technology[4] presents an approach for high-bandwidth, low-$V_\pi$ modulation that builds on silicon photonics without front-end-of-line modifications, utilizing organic electro-optic (OEO) materials that can have > 10x the electro-optic response of TFLN. While first reported 20 years ago,[5] recent advances in modulation efficiency,[6] loss,[7] and thermal stability[8, 9] have made the technology increasingly competitive, most demonstrations have been at the single-modulator level. We present multi-channel transmitter PICs (1.6T-DR8 and 3.2T-DR8) utilizing SOH modulators fabricated on commercially produced 200 mm silicon photonics wafers using a commercial crosslinkable (thermoset) OEO material and operating in the O-band with leading bandwidths and modulation efficiencies.

## PIC Fabrication

Both 1.6T and 3.2T DR8 PICs were designed by Enosemi in collaboration with NLM Photonics. PICs were fabricated on Advanced Micro Foundry (AMF)'s O-band GP v4.5 platform on a dedicated 200 mm wafer run. Each PIC includes 8 differential-drive (GSGSG) Mach-Zehnder modulators (MZMs), each with monitor photodiodes and thermal tuners for bias control. Two laser inputs with a maximum input power of +18 dBm are used, each driving four modulators. The production edge couplers are supplemented with grating coupler taps for testing. A schematic of the PIC and modulators is shown in Figure 1.

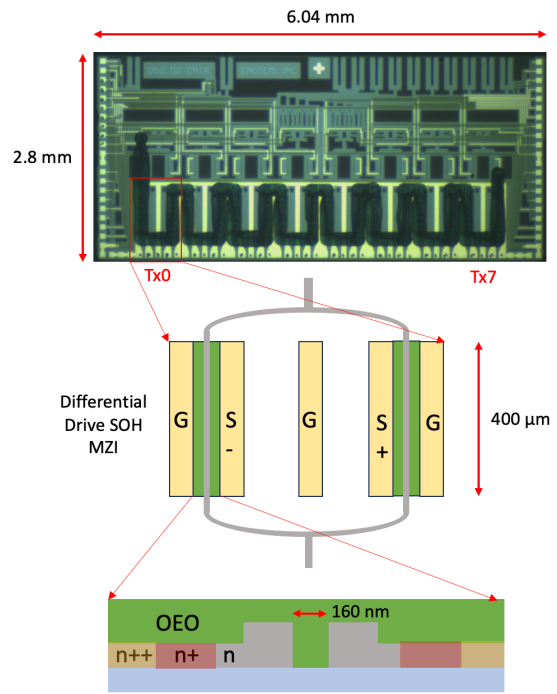

**Fig. 1** PIC and modulator schematic showing layout with locally deposited OEO material, modulator drive configuration, and slot waveguide.

Slot waveguide SOH modulators for the 1.6T DR8 PIC were designed with a phase shifter length of 400 µm, which was reduced for the 3.2T DR8 PIC; both designs used a slot width of 160 nm, rail width of 240 nm and PDK standard rib and slab thicknesses. Multi-level n-doping of silicon strip electrodes was used to optimize conductivity and bandwidth, with no gate bias required for operation. Modulators were connected to a standard multi-layer metal stack and terminated on-die for 100Ω differential drive. The compact modulators enable a total area of only 17 mm² for the 8-channel PIC.

NLM Photonics performed SOH processing at the die level, including deposition of Selerion-HTX™ (HLD) OEO material,[10] encapsulation, poling, and crosslinking for thermal stability. Due to the presence of suspended edge couplers on the AMF GP platform, blanket coating methods such as spin-coating could not be used for OEO deposition without risking damage to the edge couplers, and a custom semi-automated micro-capillary deposition system was used for OEO delivery for prototype PICs. Several alternatives have since been identified for high-volume manufacturing.

PICs were encapsulated using a conformal ALD-based encapsulation technique; all channels were then poled and crosslinked simultaneously using a multi-contact probe-based approach on a custom semi-automated probe station from Maple Leaf Photonics. While a reduced poling temperature of 115°C was used for these prototypes, Selerion-HTX can be poled at higher temperatures to achieve long-term stability at ≥ 120°C.[8, 11]

**Modulator Characterization**
Low-frequency characterization of modulators to determine bias point and $V_\pi$ was performed at NLM Photonics using a Maple Leaf probe station, using the low-frequency overmodulation method[12] for $V_\pi$ and a 100 kHz differential triangle wave as the stimulus. Characterization was performed using grating couplers.

A selection of PICs was tested at VLC photonics using a 2-port Keysight N4372 110 GHz Lightwave component analyser to obtain S-parameters and 3 dB bandwidth for multiple channels. A summary of $V_\pi$ and bandwidth results for PICs used for bandwidth or link testing is shown in Table 1.

**Table 1.** Summary of modulator performance

| PIC | $V_{\pi,diff}$ (V) | 3 dB BW (GHz) |
|---|---|---|
| 200G-A | 2.13 ± 0.12 | Not measured |
| 200G-B | 2.33 ± 0.15 | 84.7 ± 3.1 |
| 200G-C | 2.46 ± 0.63 | 84.6 ± 2.1 |
| 400G-A | 6.74 ± 0.40 | 109 ± 1 |
| 400G-B | 7.5 ± 1.8 | > 110 |

Modulators on the 200G/λ PIC were designed for a 3 dB bandwidth of 80.3 GHz, differential $V_\pi$ of 2.08V, and insertion loss of 2.4 dB at 1310 nm. Static extinction ratios were typically ≥ 25 dB with some channels > 40 dB. The best $V_{\pi,diff}$ obtained on a single channel was 1.57V. As only half the voltage swing is dropped across each arm, this yields a single-ended push-pull (SEPP) equivalent modulation efficiency of 0.31 V-mm.

The conversion enables comparison with other SOH modulators, which typically use a GSG push-pull configuration. Using the Pockels effect relationship[12] for a push-pull MZM

$$V_\pi L = \frac{\lambda_0}{2} \cdot \frac{w_{slot}}{\Gamma} \cdot \frac{1}{n_e^3 r_{33}} \quad (1)$$

with $n$ = 1.9, mode confinement factor $\Gamma$ = 0.14, and measurement wavelength $\lambda_0$ = 1290 nm gives an in-device $r_{33}$ of 343 pm/V. More consistent results could be obtained around the design target with ~10% inter-channel variation and SEPP equivalent modulation efficiencies ~0.5 V-mm ($r_{33}$ of 200-250 pm/V). Due to their shorter phase shifter length, 400G modulators had a higher design $V_{\pi,diff}$ of 5.52V but similar modulation efficiencies of ~0.5 V-mm. 3 dB EO bandwidths consistently exceeded 80 GHz for the 200G parts and were ~110GHz for the 400G parts. Representative $S_{21}$ traces are shown in Figure 2. The notch observed at approximately half the design bandwidth is a measurement artifact from performing 2-port measurements on a single arm of the modulator due to instrument limitations.

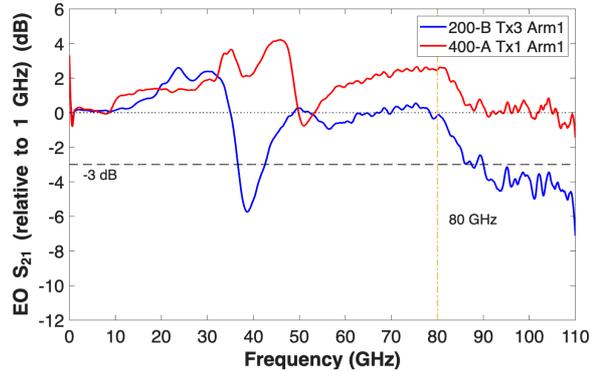

**Fig. 2** Electro-optic S21 bandwidth characterization of 200G and 400G modulators

**Link Testing**
NLM Photonics performed data transmission testing on multiple channels on two 1.6T DR8 PICs in collaboration with Keysight. PICs were tested on a Maple Leaf semi-automated probe station using a 67 GHz probe and 1.85 mm RF cabling, a Keysight M8050 BERT with M8042A pattern generator, M8009 clock, and M8059A remote head to generate the signal, and a Keysight N1000 DCA-X digital sampling oscilloscope (DSO) with N1032A optical front-end as the receiver. Illumination at 1310 nm was provided using a Santec TSL-570 tunable fibre laser. PICs were edge coupled. Modulators were biased to their quadrature point using thermal tuning before testing each channel. Due to lower laser power and reduced edge coupling efficiency without index-matching epoxy, a praseodymium-doped fibre amplifier (PDFA, Thorlabs PDFA-100) was used to restore the signal level to +4-5 dBm. Experimental diagram is shown in Figure 3.

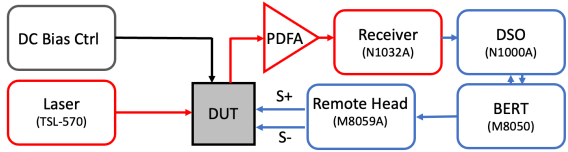

**Fig. 3** Experimental diagram for data transmission experiment

224G (112 Gbaud PAM4) data transmission tests were performed using a PRBS13Q test pattern, feed-forward linear equalization (FFE) with 35 taps, 2-level continuous time linear equalization (CTLE) and a TDECQ equalizer and measurement module. Eye diagrams for driver swings from 1.0 to 1.8 $V_{ppd}$ for Tx0 on PIC sample 200G-A are shown in Figure 4, and dynamic extinction ratios and optical modulation amplitude (OMA) in Figure 5.

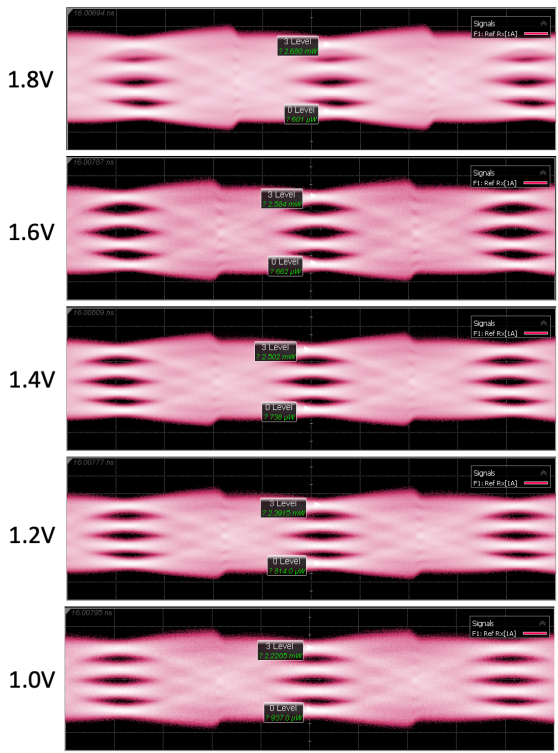

**Fig. 4** 224G PAM4 eye diagrams as a function of driver swing

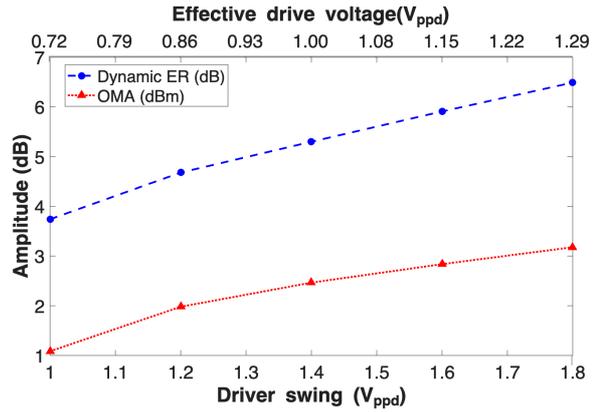

**Fig. 5** OMA and dynamic ER as a function of driver swing

PAM4 symbol error rates (SER) at 112 Gbaud were typically ~1x10$^{-2}$, below the SD-FEC threshold. A summary of link data is shown in Table 2.

**Table 2.** Summary of 224G link data

| Channel | Driver swing ($V_{ppd}$) | ER (dB) | SER |
|---|---|---|---|
| 200-B-Tx2 | 1.8 | 6.79 | 1.17x10$^{-2}$ |
| 200-B-Tx3 | 1.8 | 2.69 | 9.9x10$^{-3}$ |
| 200-A-Tx3 | 1.8 | 4.43 | 9.4x10$^{-3}$ |
| 200-A-Tx0 | 1.8 | 6.49 | 1.69x10$^{-2}$ |
| 200-A-Tx0 | 1.6 | 5.91 | 1.06x10$^{-2}$ |
| 200-A-Tx0 | 1.4 | 5.30 | 1.09x10$^{-2}$ |
| 200-A-Tx0 | 1.2 | 4.68 | 9.9x10$^{-3}$ |
| 200--Tx0 | 1.0 | 3.74 | 1.02x10$^{-2}$ |

As SER was insensitive to driver swing and relatively constant across devices, much of the error was likely introduced optically (e.g. from the PDFA) and could be reduced with improved coupling and higher input laser power. The probe and cabling introduced 2.88 dB of RF losses at the Nyquist frequency (56 GHz), such that a 1.8 $V_{ppd}$ driver swing applies 1.29 $V_{ppd}$ to the modulator. Good performance with > 5 dB dynamic ER and low OMA losses was obtained with ≤ 1V on-die voltage (1.4$V_{ppd}$ driver swing).

## Summary

We have demonstrated practical multi-channel SOH PICs capable of 1.6T and 3.2T performance, including 224G PAM4 data transmission with low driver swings. These PICs incorporated thermoset OEO materials and show repeatable multi-channel performance with leading bandwidth and modulation efficiency and demonstrate a path towards low-$V_\pi$ 400G/λ silicon photonics without front-end-of-line modifications.

## Acknowledgements

PIC development was funded in part by Centera Photonics, and development of instrumentation and process modules were supported in part by NASA (80NSSC2PB507) and AFWERX

(FA864924P0527). The authors further thank Keysight and VLC Photonics for assistance with high-speed measurements, Maple Leaf Photonics for technical support, and Drs. Anthony Yu and Tom Baehr-Jones for useful discussion, along with their colleagues at NLM Photonics and Enosemi.